\documentclass[aps,jcp,reprint,superscriptaddress,floatfix]{revtex4}

\usepackage{amsmath,amssymb,amsfonts,amsthm,mathtools}
\usepackage{bm}
\usepackage{graphicx}
\usepackage{physics}
\usepackage{hyperref}
\usepackage{booktabs}
\usepackage{enumitem}
\usepackage{tikz}
\usepackage[normalem]{ulem}

\begin{document}

\title{
Symmetry-driven correlation patterns in one-dimensional periodic fermionic
systems: closed-form expressions and exact selection rules for correlation
functions, entanglement entropies and mutual information
}

\author{Celestino Angeli}
\email{anc@unife.it}
\affiliation{Dipartimento di Scienze Chimiche, Farmaceutiche ed Agrarie, via
Borsari 46, 44121 Ferrara, Italy}

\date{\today}
\keywords{Fermionic systems; Correlation functions; Reduced density matrix; 
Mutual information; Slater determinants; Electronic structure; Quantum correlations}

\begin{abstract}
We present an analytical study of the spatial structure of correlations in
periodic fermionic systems at the single Slater determinant level, focusing on
the cyclic case. Exploiting cyclic symmetry, the transformation from localized
orbitals to molecular orbitals is simultaneously a discrete Fourier transform,
a Vandermonde matrix on the roots of unity, a complex Hadamard matrix, and the
character table of the cyclic group $C_m$. Within this framework, the spatial
structure of correlations is fully determined by the occupied irreducible
representations (Fourier modes), independently of the specific Hamiltonian
generating them. In particular, the fermionic two-point correlator is expressed
as a truncated Fourier sum over the occupied orbitals, leading to exact
analytical expressions for symmetric occupation patterns of Fourier modes
($\ell$ and $-\ell$ pairs), corresponding in quantum chemistry to aromatic
fillings, \textit{i.e.} fillings satisfying H\"uckel's $4n + 2$ rule. For the
half filling case, we derive exact spatial selection rules. In particular, the
two-point correlator vanishes identically for all even distances along the
ring. This property propagates to reduced density matrices and implies a
complete absence of mutual information between the corresponding orbitals.
These results reveal that, despite the global delocalization of molecular
orbitals, the correlation structure in cyclic systems is highly non-uniform and
governed by symmetry-induced interference. Although formulated in terms of
cyclic lattice systems, the present results extend straightforwardly to
one-dimensional fermionic chains with periodic conditions, reflecting the
underlying translational symmetry of the problem. The framework therefore
provides a Hamiltonian-independent and universal reference for understanding
correlation patterns in one-dimensional periodic fermionic systems.
\end{abstract}

\maketitle

\section{Introduction}

The spatial structure of electronic correlations is a central problem in
quantum chemistry and many-body physics. In recent years, information-theoretic
quantities such as entanglement entropy and mutual information \cite{Vedral02}
have emerged as powerful tools to analyze correlation patterns in molecular
systems,\cite{Rissl06,Barcza11,Bogus12,Pitta15,Ding20,Tenti24,Alive24,Hende24,Mater24b,Mater25,Evang25,Zhao25,Pittn25}
providing a complementary perspective to traditional approaches based on
energies and wavefunction amplitudes. More generally, entanglement has been
recognized as a fundamental organizing principle of many-electron
wavefunctions, determining the complexity of their representation in the
Hilbert space and motivating tensor-network approaches to strongly correlated
systems.\cite{Chan12} Although entanglement entropy and mutual information are
among the most widely used quantities to investigate many-body systems (even if
they account for both quantum and classical correlation
\cite{Legeza06,Modi10,Ding20,Ding21,Alive24,Mater24}), various measures have
been developed to capture different aspects of quantum
correlations.\cite{Horod09} For bipartite systems, one can recall
concurrence,\cite{Woott98} negativity,\cite{Vidal02} and quantum
discord,\cite{Olliv01,Hende01} each capturing different aspects of quantum
correlations.  Recent developments have also focused on symmetry-resolved
formulations of entanglement measures in free-fermion systems, where the
contribution of different conserved-charge sectors can be analyzed
separately,\cite{Jones22} or have emphasized that suitable orbital or qubit
transformations can substantially alter the entanglement structure of many-body
wavefunctions,\cite{Tenti24} reducing their complexity and improving
tensor-network representations.\cite{Chan23} In addition to bipartite
correlation measures, multipartite entanglement has emerged as a central
concept in the description of many-body quantum systems, with the seminal
Coffman--Kundu--Wootters monogamy relation providing one of the earliest
quantitative frameworks for its characterization \cite{Coffm00} (see Refs.
\cite{Pezze17,Troiani25,Troiani25b} for some examples of recent studies). 

Among these quantities, mutual information is particularly attractive because
it captures the total amount of correlations, both classical and quantum, and
can be directly computed from reduced density matrices. In quantum chemistry,
mutual information and related orbital-entanglement
measures have been successfully employed to analyze orbital correlation
patterns, chemical bonding, and bond-formation processes from
reduced density matrices of correlated
wavefunctions.\cite{Bogus13,Bogus15,Duper15,Szalay17,Ding26} Previous studies
have shown that mutual information can reveal nontrivial spatial correlation
patterns in free-fermion systems and encode information on subsystem geometry
and separation.\cite{Lepori22} For these reasons, in the present work, we focus
on entanglement entropies and mutual information, which provide a natural
framework for analyzing the structure of correlations in cyclic and periodic
fermionic systems. Moreover, since these quantities (and the information they
can provide about correlation) depend strongly on the partition of the system,
we will focus our attention on the case of partitions involving local sites.

Given that the exact solution of the Schr\"odinger equation is possible only
for very simple or prototypical cases, to study real systems of interest, one
must resort to appropriate approximations. On the one hand, one can try to
approximate  the wavefunction, for example, by imposing a well-defined
structure (single Slater determinant, Coupled Cluster ansatz, etc.) or
exploiting the variational method or building it step by step in a perturbative
logic. At the same time, one can simplify the Hamiltonian by introducing model
Hamiltonians that aim to capture the essential aspects of the complete
Hamiltonian. 

Systems requiring a quantum mechanical treatment are typically grouped into
different categories (such as, for instance, molecules, solid state, fermion
gas, etc.), and this categorization has led to the development of different
disciplines (\textit{e.g.}, quantum chemistry, QC, many-body theory, MBT,
etc.), each characterized by its own jargon and specific approximations,
optimal for the studied systems. In various cases, different disciplines have
identified similar approximations, which, however, are referred to by different
names. Regarding the identification of simplified model  Hamiltonians, relevant
examples for this work are the H\"uckel model Hamiltonian (HMH),\cite{Huckel31}
originally introduced in QC for the study of $\pi$ electrons in conjugated
polyenes, and the tight-binding model Hamiltonian (TBMH), defined in condensed
matter/many-body physics to describe problems such as electrons in crystal
lattices. In both models there is one parameter, the resonance integral $\beta$
in the HMH and the hopping parameter (amplitude) $t$ in the TBMH and the two
parameters differ in sign: $\beta=-t$. For this reason, hereafter HMH and TBMH
are considered synonymous.

Another example relevant to this work and involving the structure of the wavefunction
is the monodeterminant Hartree-Fock (HF) approximation in QC and the
concept of particle-number-conserving  fermionic Gaussian states in MBT. In
both cases, we are dealing with a mean-field (or uncorrelated) description of a
system composed of fermions. In the first case, the emphasis is on the
monodeterminant nature of the wavefunction, obtained by defining suitable
one-electron functions (molecular orbitals, MOs) and constructing the Slater
determinant by occupying the lowest-energy MOs (Aufbau principle), while in the
second case, the emphasis is on the correlation properties between sites. In
summary, the HF ansatz can be seen as a quantum chemical realization of
fermionic Gaussian states widely used in MBT.

Starting from these considerations for a generic fermionic system, we show in
this work that for a class of cyclic systems (those for which a single Slater
determinant is a reasonable approximation), highly nontrivial and universal
spatial organization of correlations are observed. As we shall show, these
correlation patterns are not tied to a particular Hamiltonian but emerge from
the symmetry of the system and the occupation of its symmetry-adapted MOs. In
passing, one can note that the lattice topology of a $n$-site ring is
isomorphic to that of a one-dimensional periodic chain, so the results obtained
here can be straightforwardly extended to this case.  In the systems here
considered, the number $n$ of electrons satisfies the constraint $n=4i+2$
($i\in\mathbb{N}_0$), which in chemistry is known as the H\"uckel
rule,\cite{Huckel31} one of the requirements for cyclic conjugated polyenes to
be ``aromatic'' (for these molecules the sites are carbon atoms), a key concept
in this discipline. In the following, we indicate with aromatic filling (or
symmetric filling) this condition for $n$. Particular attention will be paid to
the case where the number of sites $m$ is equal to the number of electrons
(also known as the half filling condition).

We show that the transformation from localized orbitals to molecular orbitals,
which in the group theory language is defined by the character table of the
$C_m$ symmetry group, is carried out by a matrix that simultaneously admits
multiple interpretations: a Vandermonde matrix
\cite{Ycart2013,HornJohnson1991,GolubVanLoan2013} over roots of unity,  a
discrete Fourier transform \cite{oppenheim2010discrete} (DFT), and a complex
Hadamard matrix.\cite{Turyn1970,TadejZyczkowski2006} This unified structure
provides a transparent interpretation of molecular orbitals as symmetry-adapted
coherent superpositions of localized states.

Adopting both the quantum-chemical and many-body perspectives, the goal of this
work is to develop a unified and fully analytical description of these systems
at the level of single Slater determinants (fermionic Gaussian states), without
reference to a specific Hamiltonian. We show that the spatial structure of
correlations is entirely determined by the symmetry of the system and by the
occupation pattern of the one-particle states, rather than by the details of
the underlying electronic Hamiltonian.  Within this framework, we derive the
exact analytical expressions for the fermionic two-point correlator (or
correlation matrix), with symmetric occupation patterns (corresponding to
aromatic fillings), as a truncated sum over the occupied Fourier modes (or
equivalently, over the characters of the irreducible representations of $C_m$
for the occupied orbitals).  In the special case of half filling, this leads to
exact spatial selection rules.  A central result of this work is that these
selection rules propagate to all correlation measures derived from the reduced
density matrix. In particular, we demonstrate that the mutual information
between pairs of localized orbitals/sites obeys exact spatial constraints,
leading to strictly vanishing correlations at specific distances.  To the best
of our knowledge, this provides one of the first explicit analytical
demonstrations of exact zeros of mutual information in fermionic systems at the
level of a single Slater determinant.

These results provide a new perspective on the electronic structure in these
systems (conjugated cyclic hydrocarbons, periodic chains, hydrogen rings, etc).
While molecular (or equivalently Bloch \cite{Bloch29} or Fourier transform)
orbitals are fully delocalized, the correlation structure in the localized
(atomic, Wannier \cite{Wannier37,Mazza97} or site) basis is highly non-uniform
and governed by symmetry-induced interference.  From the perspective of QC,
these results provide a new way to interpret the structure of electronic
correlations in terms of interference and symmetry, complementing the usual
picture based on orbital occupations and configurations. From the perspective
of MBT, they offer an explicit realization of how Gaussian states can exhibit
nontrivial spatial entanglement patterns purely as a consequence of symmetry
and phase coherence.  The framework developed here thus establishes a direct
and transparent connection between reduced density matrix theory, symmetry
group theory, and information theory measures of correlation. 

The paper is organized as follows. In Sec.~\ref{sec:rdm-ker} we discuss the
relation between the correlation matrix and the one-particle reduced density
matrix. In Sec.~\ref{sec:mos} we introduce the symmetry-adapted orbital basis
in cyclic systems. In Sec.~\ref{sec:VHFstr} we analyze the
Vandermonde--Hadamard--Fourier structure of the transformation matrix. In
Sec.~\ref{sec:kernel} we derive the analytical form of the elements of the
correlation matrix, while in Sec.~\ref{sec:rdm-e-e} the analytical derivation
is extended to entropies and mutual information. The parity selection rule and
some asymptotic behavior are discussed in Sec.~\ref{sec:PSR}. In
Sec.~\ref{sec:chem-inter} the analysis derived in the previous sections is
considered from a chemical point of view, deriving the main quantities for the
case of real molecular systems (neutral for the half filling case and charged
when this constraint is removed).  Finally, in Sec.~\ref{sec:concl} some
general conclusions are reported.

\section{One-particle density matrix and Slater determinants\label{sec:rdm-ker}}

The description of the electronic structure in QC and MBT is based on the concept of 
reduced density matrices and correlation
functions. In order to establish a common language, we begin by making explicit
the correspondence between the one-particle reduced density matrix (1-RDM) and
the correlation matrix. 
In quantum chemistry, electronic correlations are naturally described in terms
of reduced density matrices.\cite{Lowdin55,coleman2000reduced} In particular, the 1-RDM is
\begin{equation}
\gamma_{ij} = \langle{ a_i^\dagger a_j}{\rangle} 
\label{eq:1-rdm}
\end{equation}
where $\{a_i^\dagger, a_j\}$ are fermionic creation and annihilation operators
associated with a chosen single-particle basis, typically MOs.  The matrix
${\gamma}$ encodes both orbital occupations (through its diagonal elements) and
delocalization or coherence between orbitals (through its off-diagonal
elements), while higher-order reduced density matrices capture genuine
many-body correlations beyond mean-field descriptions. 

In parallel, fermionic systems are characterized in MBT by correlation
functions,\cite{peschel2003,Pesch2009} with the correlation matrix $K$ with
elements
\begin{equation}
K_{ij} = {\langle}{ c_i^\dagger c_j}{\rangle}, 
\label{eq:kernel}
\end{equation}
playing a fundamental role. 
Here, $\{c_i^\dagger, c_j\}$ are fermionic operators on a site or lattice
basis. 

Although the two equations are formally identical, it is important to emphasize
that in general they apply to different contexts. In quantum chemistry,
operators $a^\dagger$ ($a$) in Eq. \ref{eq:1-rdm} are related to MOs,
one-particle functions potentially delocalized throughout the molecule (they
give a priority to the kinetic energy \cite{Angeli24}), while in Eq.
\ref{eq:kernel} it is usually assumed that $c^\dagger$ and $c$ are associated
to localized functions (site, Wannier functions,\cite{Wannier37,Mazza97} etc.),
as is also evident from the fact that ${K}$ is also known as two-point
correlator.  When ${\gamma}$ is also expressed on a localized basis, they
coincide, providing a direct bridge between the two frameworks.

\subsection{Slater determinants and Gaussian states\label{sec:gaussian}}

A particularly important approximation for the ground state wavefunction of
many molecular systems is given by a single Slater determinant, used in popular
approaches such as, for instance, the Hartree--Fock (HF) method as well as the
TBMH and HMH.  From a many-body perspective, these states are fermionic
Gaussian states, meaning that all correlation functions and reduced density
matrices are completely determined by the correlation matrix via Wick's
theorem.\cite{Wick50,Abrik63,Fetter71,Mahan00,Altla2010} As a consequence, the
full correlation and entanglement structure of these systems is encoded in a
single object, the correlation matrix.  This property is well known in both
communities, although it is often formulated in different
terms.\cite{szabo_ostlund,peschel2003}

When studying molecules, the one-particle functions (MOs) used in the
monodeterminant approximation are typically ``delocalized'' across multiple
sites/atoms, that is, they are linear combinations of local functions.
Inserting this linear combination into the Slater determinant yields a linear
combination of Slater determinants expressed on local functions, with
coefficients defined by the coefficients of the MO expansions on the local
functions. In quantum chemistry, when the local basis is chosen to be
orthonormal, this rewriting is called the orthogonal valence bond (OVB) reading
of the wavefunction.\cite{Genesis,JCE-OVB,MP-OVB,Angeli24}

Let $\{ \phi_\ell \}$ denote a set of MOs, and
let the many-body state $\Psi$ be constructed by occupying a subset of them:
\begin{equation}
|\Psi\rangle = \left(\prod_{\ell \in \mathrm{occ}} a_\ell^\dagger\right) |0\rangle,
\end{equation}
where $a_\ell^\dagger$ creates an electron in orbital $\phi_\ell$. The
operators $a_\ell^\dagger$ are related to the operators $c_i^\dagger$ on the 
localized orthonormal basis through a unitary transformation:
\begin{equation}
a_\ell^\dagger = \sum_i U_{i\ell} \, c_i^\dagger.
\end{equation}

In this framework, the elements $K_{ij}$ of the correlation matrix take the form
\begin{equation}
K_{ij} = {\langle}{c_i^\dagger c_j}{\rangle} = \sum_{\ell \in \mathrm{occ}} U_{i\ell}^* U_{j\ell}.
\label{eq:one-body-ker}
\end{equation}
This expression shows that the correlation matrix is a projector onto the subspace of
occupied orbitals, expressed in the local basis
\begin{equation}
K = U_{\mathrm{occ}} U_{\mathrm{occ}}^\dagger,
\end{equation}
where $U_{\mathrm{occ}}$ denotes the matrix formed by the columns of $U$
corresponding to the occupied MOs.

\subsection{Higher-order correlations and Wick factorization}

A relevant property of Slater determinants is that all higher-order correlation
functions are determined entirely by the correlation matrix. In quantum chemistry,
this is expressed, for instance, by the fact that the two-particle reduced density matrix
(2-RDM) can be written in terms of the 1-RDM as
\begin{equation}
\Gamma_{ijkl} = {\langle}{a_i^\dagger a_j^\dagger a_l a_k}{\rangle}
= \gamma_{ik}\gamma_{jl} - \gamma_{il}\gamma_{jk}.
\end{equation}
In many-body physics, all higher-order correlators are expressed as a sum over
the products of elements of the correlation matrix. For instance, the
four-point correlation function is
\begin{equation}
{\langle}{c_i^\dagger c_j^\dagger c_k c_l}{\rangle}=
{\langle}{c_i^\dagger c_l}{\rangle} {\langle}{c_j^\dagger c_k}{\rangle}-
{\langle}{c_i^\dagger c_k}{\rangle} {\langle}{c_j^\dagger c_l}{\rangle}=
K_{il}K_{jk}-K_{ik}K_{jl}.
\end{equation}
As a consequence, Slater determinants do not contain genuine many-body
correlations beyond those induced by antisymmetry. All physical observables,
including reduced density matrices and correlation measures, are fully
determined by the correlation matrix.

The fact that the entire structure of a Slater determinant is encoded in the
correlation matrix has important implications for correlation and entanglement.
In particular, the reduced density matrix of any subsystem, and therefore
quantities such as entanglement entropy and mutual information, can be
expressed solely in terms of the restriction of $K_{ij}$ to that subsystem.

\section{Cyclic symmetry and molecular orbitals\label{sec:mos}}

We now consider systems with cyclic symmetry, such as conjugated hydrocarbon
rings, which are invariant under discrete rotations. If $m$ equivalent sites
are symmetrically (with equal spacing) arranged on a circle, the symmetry group
is the cyclic group $C_m$ generated by the counterclockwise rotation through an
angle $2\pi/m$, indicated with $C_m$ (using the same symbol to indicate the
group and the operation that generates it can be confusing, but this is the
usual usage). Actually, $C_m$ is a subgroup of the full symmetry group of the
system (which also contains, for instance, symmetry planes), but staying within
this subgroup is an effective choice for the study reported hereafter. The $m$
symmetry elements of the group are indicated with $C_m^j$, with $j$ indicating
the number of times the $C_m$ operation is applied ($C_m^0$ indicates the
identity), $0\leq j \leq m-1$.

\subsection{Symmetry-adapted orbitals \label{sec:salc}}

Let $\{ | j \rangle \}$ denote a set of localized orbitals (for instance, carbon
atomic $2p_z$ orbitals in $\pi$-conjugated cyclic polyenes, or hydrogen atomic
$1s$ orbitals in hydrogen rings), labeled by their position $j = 0, \dots,
m-1$. If these localized orbitals are  numbered in a counterclockwise
direction,  $C_m$ maps orbital $j$ to orbital $j+1$ modulo $m$. This symmetry
plays a central role in determining the structure of the single-particle
states and, consequently, of the many-electron wavefunction.

Given that the cyclic group $C_m$ is abelian, all its irreducible
representations (irreps) are one-dimensional and can be labeled by an integer
$\ell = 0, \dots, m-1$. The character of the element $C_m^j$ in the $\ell-th$
irrep is given \cite{hamermesh1989group,cotton1990chemical} by
\begin{equation}
\chi^\ell(C_m^j) = \omega^{\ell j}, \qquad \omega = e^{i\frac{2\pi}{m}}.
\label{eq:char-tab}
\end{equation}

Symmetry-adapted molecular orbitals are obtained by projecting the localized
basis onto these irreps, leading to the well-known expression
\begin{equation}
| \phi_\ell \rangle = \frac{1}{\sqrt{m}} \sum_{j=0}^{m-1} \omega^{\ell j} | j \rangle,
\label{eq:salc}
\end{equation}
which defines a complete orthonormal set of delocalized orbitals.  From a
quantum chemical perspective, these are precisely the H\"uckel or the canonical
HF (regardless of computational details, such as the atomic basis set) MOs for
a cyclic system. From a many-body perspective, they correspond to plane-wave
(momentum) eigenstates on a discrete ring.

It is important to highlight that, belonging to different irreps, MOs with this
structure are not only the eigenstates of the HMH (or TBMH) but more generally
of any totally symmetric one-electron Hamiltonian, such as the Fock operator of
HF theory (or its generalizations in Multi-Reference approaches).  Obviously,
while in the HMH/TBMH case the local orbitals do not require an explicit
definition, in the case of the Fock operator (introduced within the HF
treatment and which depends on the chosen atomic basis set) the issue is
slightly more complex, but various localization techniques exist that allow one
to define a suitable set of localized (orthogonal) functions (see Refs.
\cite{Lenna49,Lenna49b,Hall50,Boys60} for some seminal works).

This consideration is the basis for  the fact that the results obtained in this
work are completely general and are not tied to a specific choice of the
Hamiltonian used.

\subsection{Fourier representation}

The transformation from the localized basis $\{ | j \rangle \}$ to the
symmetry-adapted basis $\{ | \phi_\ell \rangle \}$ reported in the previous section
can be written in matrix form as
\begin{equation}
| \phi_\ell \rangle = \sum_j U_{j\ell} | j \rangle,
\end{equation}
\begin{equation}
U_{j\ell} = \frac{1}{\sqrt{m}} \omega^{\ell j},
\label{eq:matr-U}
\end{equation}
with $\omega$ defined in Eq. \ref{eq:char-tab}.

The matrix $U$ is the DFT on a lattice with $m$ sites.  The DFT
yields lattice-momentum eigenstates, \textit{i.e.} discrete Bloch waves, which
diagonalize any single-particle Hamiltonian with periodic cyclic/translational 
symmetry, including the HMH, TBMH, and the Fock operator in HF. 

In second-quantized form, the transformation reads
\begin{equation}
a_\ell^\dagger = \frac{1}{\sqrt{m}} \sum_j \omega^{\ell j} c_j^\dagger,
\end{equation}
where $a_\ell^\dagger$ creates an electron in the delocalized orbital
$\phi_\ell$, and $c_j^\dagger$ creates an electron in the localized orbital at
the site $j$.

\subsection{Connection to continuous angular momentum\label{sec:moment}}

The index $\ell$ admits a natural interpretation in terms of
angular momentum. In a continuous description of electrons moving on a ring,
the single-particle states are eigenfunctions of the angular momentum operator
\begin{equation}
L_z = -i\hbar \frac{d}{d\theta},
\end{equation}
($z$ is the direction orthogonal to the ring plane)
with eigenvalues $\hbar \ell$, where $\ell \in \mathbb{Z}$.

In the presence of a potential with $C_m$ symmetry, such as that generated by
the discrete arrangement of nuclei in a cyclic system, the continuous
rotational symmetry is reduced to a discrete subgroup. As a consequence, $L_z$
is no longer conserved and the eigenstates are no longer characterized by a
definite angular momentum. However, the system retains invariance under
discrete rotations, and the relevant quantum number becomes the angular
momentum modulo $m$. 
In this setting, the index $\ell$ labeling the symmetry-adapted orbitals can be
interpreted as a discrete quantum number of the angular momentum, corresponding to the
projection of the continuous angular momentum $L_z/\hbar$ modulo $m$.
Equivalently, $\ell$ labels the irreducible representations of the cyclic group
$C_m$ and determines the phase acquired under rotation by $2\pi/m$.

This interpretation provides a direct connection between the discrete Fourier
structure of the molecular orbitals and the angular momentum description of
electrons on a ring and clarifies the origin of the pairing $\ell
\leftrightarrow -\ell$, which reflects the invariance under time-reversal and
the reality of physical observables. 

\subsection{Occupation patterns and symmetry\label{sec:aufbau}}

It is well known that the eigenvalues of TBMH for a cyclic system are
\begin{equation}
\varepsilon_l=-2t\cos\left(\frac{2\pi l}{m}\right)
\label{eq:epsilon}
\end{equation}
and one can note that $\varepsilon_0=-2t$ is the lowest (non degenerate) value,
while the eigenvalues corresponding to $\ell$ and $m-\ell$ (or $-\ell$, everything
is defined modulo $m$) with $\ell\ne 0$ are degenerate.

The $\ell \leftrightarrow -\ell$ degeneracy is not specific to the TBMH
approximation but holds generally for periodic one-dimensional Hamiltonians
invariant under time-reversal (or momentum inversion). From a different point
of view, one can note that in addition to the symmetry elements present in the
$C_m$ group, the full symmetry group contains other elements, for instance
symmetry planes, which are orthogonal to the plane of the ring and contain the
axis orthogonal to it and passing through its center. Since these planes
commute with all one-electron Hamiltonians respecting the system symmetry and
they transform $| \phi_\ell \rangle$ into  $|\phi_{-\ell} \rangle$ and
\textit{vice versa}, these two orbitals are degenerate. For these reasons,
$\ell \leftrightarrow -\ell$ degeneracy is assumed to be general hereafter.

In a Slater determinant, the many-electron state is specified by the set of
occupied orbitals. For the ground state, they are chosen on an energetic basis
following the Aufbau principle.  In the present work, however, the essential
assumption is not the detailed energetic origin of the orbitals, but the
existence of a symmetry-preserving single-determinant reference with a compact
and symmetric occupation of the Fourier modes.

A particularly important situation arises when the occupied orbitals
are symmetrically distributed in the index $\ell$.  In other words, if an orbital
with index $\ell$ is occupied with two electrons, the orbital with index
$-\ell$ is also occupied with two electrons. 
This symmetric occupation pattern identifies the natural symmetry-preserving 
single-determinant reference. It may provide
a good approximation to the exact wavefunction in the H\"uckel/tight-binding models
and in weakly correlated closed-shell regimes, including small $|U|$ 
Hubbard type descriptions or mean-field treatments of cyclic polyenes
based on the full electronic Hamiltonian. On the other hand, it is not intended to describe strongly correlated
limits, such as Hubbard rings at large positive or negative $U$, where qualitatively different
many-body states may emerge. In this sense, the results derived below should be understood as 
exact results within the Slater-determinant/Gaussian reference. Deviations from these
symmetry-induced patterns in correlated calculations provide signatures of genuine 
many-body correlation.

The symmetric occupation
condition is fulfilled when the number of electrons $n$ satisfies the
well-known H\"uckel rule $n=4\ell_{max}+2$ ($\ell_{max}\in\mathbb{N}_0$ is the
maximum value in modulus of the values of $\ell$ for the occupied orbitals).
This occupation pattern (for spinless
fermions or within a fixed spin sector) ensures that the correlation matrix is
real-valued, as contributions from $\ell$ and $-\ell$ combine into cosine terms
(see hereafter Sec.~\ref{sec:kernel}). 

As we show in the following sections, this structure implies that the elements
$K_{ij}$ in real space are given by a superposition of phase factors associated
with the occupied representations. The resulting interference effects are at
the origin of the spatial organization of correlations and lead to exact
selection rules that are independent of the details of the Hamiltonian.

\section{Vandermonde--Hadamard--Fourier structure\label{sec:VHFstr}}

The transformation matrix $U$ introduced in Eqs. \ref{eq:char-tab} and
\ref{eq:matr-U} plays a central role in the structure of cyclic systems. While
it is commonly recognized as the DFT, it admits several equivalent
interpretations that provide complementary insights into the origin of the
correlation structure. We have already seen (see Sec. \ref{sec:salc}) that it
also encodes the character table of the $C_m$ group. Here we show two other
interpretations.

\paragraph{Vandermonde structure}

Let us define the set of complex numbers
\begin{equation}
x_j = \omega^j, \qquad j = 0, \dots, m-1,
\end{equation}
($\omega$ defined in Eq. \ref{eq:char-tab})
which correspond to the $m$-th roots of unity distributed uniformly on the unit
circle in the complex plane. In terms of these variables, the matrix $U$ can be written as
\begin{equation}
U_{j\ell} = \frac{1}{\sqrt{m}} x_j^\ell.
\end{equation}
Up to normalization, this is a Vandermonde matrix \cite{Ycart2013,HornJohnson1991,GolubVanLoan2013} 
constructed from the points $\{x_j\}$. This representation emphasizes the
algebraic structure of the transformation and connects it to polynomial
interpolation on the unit circle.

\paragraph{Complex Hadamard matrix}

Another important property of $U$ is that it is unitary and has elements of equal modulus:
\begin{equation}
|U_{j\ell}| = \frac{1}{\sqrt{m}}, \qquad U U^\dagger = I.
\end{equation}
Matrices with these properties are known as complex Hadamard matrices.\cite{Turyn1970,TadejZyczkowski2006} They
generalize the familiar Hadamard matrices encountered in quantum information
theory and are characterized by the fact that all rows (and columns) are
mutually orthogonal and have equal norm.

From a physical perspective, this implies that each delocalized orbital $|\phi_\ell\rangle$ is
spread uniformly over all sites, differing only by phase factors. Therefore, for all orbital and for all site $j$ the weight  is the same, given that $|e^{i\frac{2\pi l j}{m}}/\sqrt{m}|^2=1/m$ Conversely,
each localized orbital is an equal-weight superposition of all Fourier modes.
The transformation is therefore maximally delocalizing and preserves the total
probability amplitude uniformly across the system.

It is worth noting that matrices of this type play a central role in quantum
information theory, where they are used to construct uniform superpositions and
to implement Fourier transforms in quantum algorithms. In this context, the
Hadamard matrix (for $m=2$) and its generalizations are used to generate states
with maximal coherence across basis states.

The appearance of the same structure in cyclic molecular systems suggests a
close analogy: the delocalized molecular orbitals can be viewed as coherent
superpositions generated by a Fourier-type transformation and the resulting
correlation patterns arise from the interference of these phases. While we will
not pursue this connection further here, it points to a possible link between
the structure of electronic states in molecules and the role of Fourier
transforms in quantum information processing.

\section{Correlation matrix as a truncated Fourier transform\label{sec:kernel}}

For a Slater determinant constructed from a set of occupied orbitals $\{
\phi_\ell \}$, the correlation matrix on the localized basis is given by
Eq.~\ref{eq:one-body-ker}. Since we are considering a closed-shell Slater
determinant, the $\alpha$ and $\beta$ spin sectors are uncorrelated and we can
limit our attention to a single sector.  Therefore, what follows will concern
only one spin sector with $N=n/2$ electrons.

Using the explicit form of the transformation matrix reported in Eqs. \ref{eq:char-tab} and \ref{eq:matr-U}
we obtain
\begin{equation}
K_{ij} = \frac{1}{m} \sum_{\ell \in \mathrm{occ}} \omega^{\ell (j-i)}.
\end{equation}

Introducing the distance $d = j - i$ modulo $m$, the elements of the
correlation matrix depend only on $d$ and one can introduce the
rotation-invariant two-point correlation function $K(d)$,
\begin{equation}
K(d)\equiv K_{ij}  = \frac{1}{m} \sum_{\ell \in \mathrm{occ}} \omega^{\ell d}.
\end{equation}

For a contiguous block of occupied orbitals symmetric around $\ell=0$, which is
the case for the ground states described in Sec.~~\ref{sec:aufbau},
the sum can be analytically evaluated. Following the notation of Sec.~~\ref{sec:aufbau}, the
occupied set is $\ell = -\ell_{max}, \dots, \ell_{max}$ (modulo $m$). Then
\begin{equation}
K(d) = \frac{1}{m} \sum_{\ell=-\ell_{max}}^{\ell_{max}} e^{i\frac{2\pi  \ell d }{m}}.
\label{eq:kd}
\end{equation}
Note that for $d=0$, one has
\begin{equation}
K(0)\equiv K_{ii} = \frac{N}{m} \quad\forall i=0,1,\cdots,m-1
\label{eq:k0}
\end{equation}
which is the occupation of a site and which, as expected from the
rotational/translational symmetry, is the same for all sites.  The sum in
Eq.~\ref{eq:kd} can be evaluated as a finite geometric series, eventually
yielding
\begin{equation}
K(d) = \frac{1}{m} \frac{\sin\left[(2\ell_{max}+1)\frac{\pi d}{m}\right]}{\sin\left(\frac{\pi d}{m}\right)}.
\end{equation}
This known expression (see, for instance, Ref. \cite{Lepetit89} and more
recently, \cite{Ding24}) provides an exact analytical form for the two-point
correlation function at all distances. Equivalent finite-ring correlation
matrices have appeared in studies of entanglement Hamiltonians for free-fermion
chains.\cite{Eisler18}

We can now recall that we are considering a filling that meets the H\"uckel rule ($n=4\ell_{max}+2$)
and therefore $2\ell_{max}+1=n/2=N$
\begin{equation}
K(d) = \frac{1}{m} \frac{\sin\left(\frac{N\pi d}{m}\right)}{\sin\left(\frac{\pi d}{m}\right)}.
\label{eq:ker-general}
\end{equation}
This formula makes explicit the interference structure of the two-point correlation function. 
The expression derived above shows that $K(d)$ is the result of a coherent
superposition of phase factors associated with the occupied orbitals. The
closed-form expression reveals that its structure is governed by the ratio of
two sine functions, reflecting the finite size of the system and the discrete
nature of the Fourier spectrum.

\subsection{Exact selection rules for half filling\label{sec:sel-rule}}

The analytical expression derived in the previous section allows us to identify,
for the half filling case, exact and universal constraints on the spatial structure of $K(d)$. 
Indeed, in the case
of half filling ($n=m$) Eq.~\ref{eq:ker-general} becomes
\begin{equation}
K(d) = \frac{1}{m} \frac{\sin\left(\frac{\pi d}{2}\right)}{\sin\left(\frac{\pi d}{m}\right)}
\label{eq:kdhf}
\end{equation}
and the numerator vanishes for all even values of $d$. 
Since the denominator is finite for $0 < d < m-1$, it follows that
$K(d) = 0$ for all even  $d$.

This result is exact and does not rely on any approximation beyond the
single-determinant description. It therefore represents a robust and
symmetry-protected feature of cyclic fermionic systems at half filling
and with a symmetric filling of the Fourier modes (note that this implies
that the number of sites must satisfy the H\"uckel rule $m=4\ell_{max}+2$, $\ell_{max}\in\mathbb{N}_0$).
As a consequence, the vanishing of the two-point correlation function at even distances is a universal
feature of all Slater determinants satisfying these conditions. This includes,
but is not limited to, H\"uckel and tight-binding descriptions,
as well as more general mean-field states (such as HF) with the same symmetry properties.

The exact cancellation of the two-point correlation function at specific
distances has direct consequences for the spatial organization of electronic
correlations.  As we show in the next section, this property extends to reduced
density matrices and information-theoretic measures of correlation. 

\section{Reduced density matrices, entropies and entanglement\label{sec:rdm-e-e}}

\subsection{Reduced density matrices from the two-point correlation function}

As a direct consequence of Wick's theorem, for single Slater determinants the
reduced density matrix of any subsystem is completely determined by the
restriction of the correlation matrix to that subsystem.  One can recall that
in entanglement-based analyses of electronic structure, reduced density
matrices are used to extract orbital entropies and orbital-pair mutual information
measures\cite{Legeza03,Bogus15} and to analyze bond-formation processes in correlated
molecular wavefunctions.\cite{Bogus13,Duper15}

We have already obtained (see Eq. \ref{eq:k0}) that the average occupation of
the site $\langle n_i\rangle=K_{ii}$ is $N/m$.

The reduced density matrix for two lattice sites \(i\) and \(j\), $\rho_{ij}$, 
written on the basis of local occupation
$\{|00\rangle,|01\rangle,|10\rangle,|11\rangle\}$
has a block-diagonal structure due to particle-number
conservation (spin is automatically conserved given that we are considering only
one spin sector), with the form
\begin{equation}
\rho_{ij}
=
\begin{pmatrix}
\langle (1-n_i)(1-n_j)\rangle & 0 & 0 & 0\\
0 & \langle (1-n_i)n_j\rangle & \langle c_j^\dagger c_i\rangle & 0\\
0 & \langle c_i^\dagger c_j\rangle & \langle n_i(1-n_j)\rangle & 0\\
0 & 0 & 0 & \langle n_i n_j\rangle
\end{pmatrix}.
\end{equation}

For Gaussian states, Wick's theorem gives
\begin{equation}
\langle n_i n_j\rangle
=
\langle n_i\rangle
\langle n_j\rangle
-
\langle c_i^\dagger c_j\rangle
\langle c_j^\dagger c_i\rangle.
\end{equation}
Using rotational/translational invariance,
$\langle n_i\rangle
=
\langle n_j\rangle
=
\nu$,
one obtains
\begin{equation}
\langle n_i n_j\rangle
=
\nu^2-K_{ij}^2.
\end{equation}
Similarly,
\begin{equation}
\langle (1-n_i)(1-n_j)\rangle
=
(1-\nu)^2-K_{ij}^2,
\end{equation}
and
\begin{equation}
\langle (1-n_i)n_j\rangle
=
\langle n_i(1-n_j)\rangle
=
\nu(1-\nu)+K_{ij}^2.
\end{equation}

The two-site reduced density matrix is, therefore,
\begin{equation}
\rho_{ij}
=
\begin{pmatrix}
(1-\nu)^2-K_{ij}^2 & 0 & 0 & 0\\
0 & \nu(1-\nu)+K_{ij}^2 & K_{ij} & 0\\
0 & K_{ij} & \nu(1-\nu)+K_{ij}^2 & 0\\
0 & 0 & 0 & \nu^2-K_{ij}^2
\end{pmatrix}.
\end{equation}
Introducing $\lambda_\pm = \nu\pm K_{ij}$, the spectrum of $\rho_{ij}$ is $\left\{ 
(1-\lambda_+)(1-\lambda_-), \lambda_+\lambda_-, \lambda_+(1-\lambda_-), (1-\lambda_+)\lambda_-\right\}$.

\subsection{Entanglement entropy and mutual information}

Using the binary entropy function
$h(x) = -x\log x-(1-x)\log(1-x)$,
the one-orbital von Neumann entropy is 
\begin{equation}
S(\rho_{i})=h(\nu)=h\left(\frac{N}{m}\right).
\end{equation}

For the two-orbital von Neumann entropy, after a few algebraic steps, one has
\begin{equation}
S(\rho_{ij})= h(\lambda_+)+h(\lambda_-)=h\left(\frac{N}{m}+K_{ij}\right)+
h\left(\frac{N}{m}-K_{ij}\right).
\end{equation}

At this point, the mutual information is
\begin{equation}
I(i,j)=S(\rho_i)+S(\rho_j)-S(\rho_{ij})
=
2h\left(\frac{N}{m}\right)-h\left(\frac{N}{m}+K_{ij}\right)-
h\left(\frac{N}{m}-K_{ij}\right).
\label{eq:mi-finale_general}
\end{equation}

At half filling, $N/m=1/2$, thus
\begin{equation}
S(\rho_{i}) = \log(2),
\end{equation}
\begin{equation}
S(\rho_{ij}) = h\left(\frac{1}{2}+K_{ij}\right)+h\left(\frac{1}{2}-K_{ij}\right)=2h\left(\frac{1}{2}+K_{ij}\right),
\end{equation}
where the second identity uses the symmetry $h(x)=h(1-x)$, and
\begin{equation}
I(i,j) = 
2\log(2)-2h\left(\frac{1}{2}+K_{ij}\right).
\label{eq:mi-finale}
\end{equation}

From this equation, one promptly obtains that
\begin{equation}
K_{ij} = 0 \quad \Longrightarrow \quad I(i,j) = 0,
\end{equation}
which holds exactly and is a consequence of 
the fact that with $K_{ij} = 0$ the reduced two-particle density matrix takes a fully factorized form,
$\rho_{ij} = \rho_i \otimes \rho_j$.
This establishes a direct and exact correspondence between the vanishing of
single-particle coherence and the absence of total correlations between the
corresponding orbitals. 

Combining this result with the parity selection rule derived in Sec.~\ref{sec:sel-rule}, we
obtain a direct and exact constraint on the spatial structure of mutual
information, that is,
$I(d) = 0$ for all even $d$.

We recall here that the relations obtained in this section concern only one
spin sector. Since the two spin sectors are uncorrelated and equivalent, the
quantities (entanglement entropy and mutual information) considering both
sectors together are simply double those reported.

Therefore, in cyclic systems with the ground state described by a single Slater
determinant with symmetric occupation at half filling, sites separated by an
even number of sites are completely uncorrelated, not only at the level of
single-particle coherence but also in terms of information-theoretic measures.
The presence of exact zeros in the mutual information highlights the existence
of strictly decoupled subsystems within an otherwise delocalized electronic
state.  This provides an explicit analytical example of exact spatial
constraints on mutual information in fermionic systems at the level of a single
Slater determinant.  The vanishing of mutual information is therefore not an
independent property, but the final manifestation of a hierarchy of
symmetry-induced constraints propagating from the one-particle correlator to
the reduced density matrices and their associated entropic measures.

\section{Parity selection rule\label{sec:PSR}}

Let us summarize the results of Sections~\ref{sec:kernel} and~\ref{sec:rdm-e-e}
for the case of half filling ($m=n$). The special form of the two-point
correlation function immediately yields that for two distinct sites $i\neq j$:
\begin{enumerate}
\item if $d=i-j$ is even, then $K(d)=0$, hence
\begin{equation}
\langle n_i n_j\rangle=\frac14,\qquad
\rho_{ij}=\frac14\,\mathbb I_4,\qquad
I(i,j)=0;
\end{equation}
\item if $d$ is odd, then
\begin{equation}
K(d)=\frac{\left(-1\right)^{\frac{d-1}{2}}}{m\sin\left(\frac{\pi d}{m}\right)}\neq 0,
\label{eq:kdodd}
\end{equation}
and therefore $I(d)\neq 0$.
\end{enumerate}

Thus, even separations are exactly uncorrelated, while odd separations carry
all the nontrivial two-site information.

\subsection{Monotonicity along odd distances and large-distance asymptotics}

Fix $n=m=4\ell_{max}+2$. For odd separations $d=1,3,5,\dots,2\ell_{max}+1$, 
Eq. \ref{eq:kdodd} shows that $\left|K(d)\right|$ decreases as $d$ increases for
$1\le d\le n/2$. Mutual information is an increasing function of
$\left|K(d)\right|$ because $h(1/2+K(d))=h(1/2-K(d))$ 
and the binary entropy is maximal at $1/2$. 
Thus, $I(d)$ decreases along odd distances in the interval $1\le d\le n/2$.

\subsection{Large-distance asymptotics}

For odd $d$ with $1\ll d\ll n$,
\begin{equation}
\left|K(d)\right|=
\frac{1}{m\sin\left(\frac{\pi d}{m}\right)}
\sim \frac{1}{\pi d}.
\end{equation}
Moreover, for small $\kappa=K(d)$,
\begin{equation}
h\!\left(\frac12+\kappa\right)=h\!\left(\frac12-\kappa\right)=\log 2 - 2\kappa^2 - \frac{4}{3}\kappa^4 + O(\kappa^6),
\end{equation}
hence
\begin{equation}
I(i,j)=4\kappa^2+\frac{8}{3}\kappa^4+O(\kappa^6).
\end{equation}

Therefore, at large odd distances,
\begin{equation}
I(i,j)\sim \frac{4}{\pi^2 d^2}.
\end{equation}

\section{Generality and chemical interpretation\label{sec:chem-inter}}

The results derived in the previous sections establish a direct and
model-independent connection between symmetry, orbital structure, and spatial
correlations in single Slater determinants. 

In cyclic conjugated hydrocarbons, a set of molecules of significant interest
in chemistry, the delocalized molecular orbitals obtained from symmetry
considerations are uniformly spread over the ring, with phases determined by
the irreducible representation index. The two-point correlation function
derived in this work provides a spatially resolved measure of how this
delocalization manifests itself in real space. For systems satisfying the
H\"uckel rule and the half filling condition, the parity selection rule implies
that, despite the global delocalization of the molecular orbitals, the
effective coherence between certain pairs of sites vanishes exactly.  This
result reveals a nontrivial internal structure within the delocalized
electronic state, in which coherence is selectively suppressed by interference
effects.  Such a structure is not immediately apparent from the delocalized
nature of the molecular orbitals but emerges naturally when analyzed in the
localized basis.

%\subsection{Relation to aromatic systems}

The specific case of fillings satisfying $n = 4\ell_{max} + 2$, which
corresponds to aromatic systems in H\"uckel theory, plays a special role in
chemistry (consider, for example, the key role played by the benzene molecule).
This observation is particularly significant because symmetric occupation
patterns are not restricted to idealized lattice models, but they are realized
by a class of experimentally accessible aromatic molecules, providing a direct
connection between the present analytical results and real chemical systems.
In this situation, the occupation pattern leads to a particularly simple
analytical form of the two-point correlation function, and the parity selection
rule becomes exact.  Although traditional criteria for aromaticity focus on
energetic, geometric, magnetic, and electronic
descriptors,\cite{Krygo14,Merino23} or electron counting rules, the present
approach highlights the role of symmetry and phase coherence in shaping the
internal structure of the electronic state. 

This viewpoint is complementary to correlation-based
descriptions of chemical bonding, in which localized-orbital and multiorbital
correlation patterns are used to characterize bonding
structures.\cite{Szalay17}

\subsection{Application to benzene and short annulenes}

To illustrate the chemical implications of the general results derived above,
we consider a series of cyclic conjugated hydrocarbons of increasing size, starting
from benzene and extending to short annulenes.

\paragraph{Benzene ($n=m=6$).}

In the H\"uckel description, the six $\pi$ electrons of benzene occupy the
three lowest-energy $\pi$-molecular orbitals, corresponding to the Fourier
indices $\ell = 0, \pm 1$. In this case, the sites are the $2p_z$ (orthogonal)
atomic orbitals of the carbon atoms (the $z$ axis is orthogonal to the benzene
plane).  The two-point correlation function  on this local basis (see Eq.
\ref{eq:kdhf}) is given by
\begin{equation}
K(d) 
= \frac{1}{6} \frac{\sin\left(\frac{\pi d}{2}\right)}{\sin\left(\frac{\pi d}{6}\right)}.
\end{equation}
This leads to the values
\begin{equation}
K(1) = \frac{1}{3}, \quad
K(2) = 0, \quad
K(3) = -\frac{1}{6}.
\end{equation}

As discussed above, the vanishing of $K(2)$ implies a complete absence of
mutual information between second-neighbor orbitals, while for $d=1$ and $d=3$,
following Eq. \ref{eq:mi-finale} one has
\begin{equation}
I(1)=2\log(2)-2h\left(\frac{1}{2}+\frac{1}{3}\right)\simeq 0.485172,
\end{equation}
\begin{equation}
I(3)=2\log(2)-2h\left(\frac{1}{2}+\frac{1}{6}\right)\simeq 0.113266.
\end{equation}
Similar results for mutual information have been reported  in Ref.
\cite{Tenti24} where, however, the higher quality wavefunction (CASSCF with 6
electrons in 6 orbitals) leads to a deviation from the exact parity selection
rules, although the values at even distances remain small.

\paragraph{Larger rings.}

\begin{figure}[h]
\centering
\includegraphics[width=0.45\linewidth]{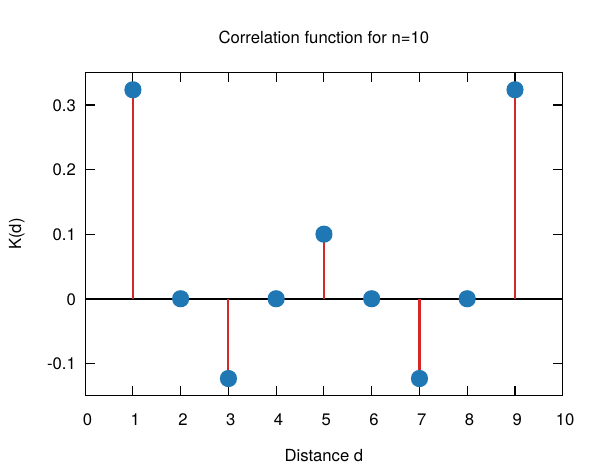}
\includegraphics[width=0.45\linewidth]{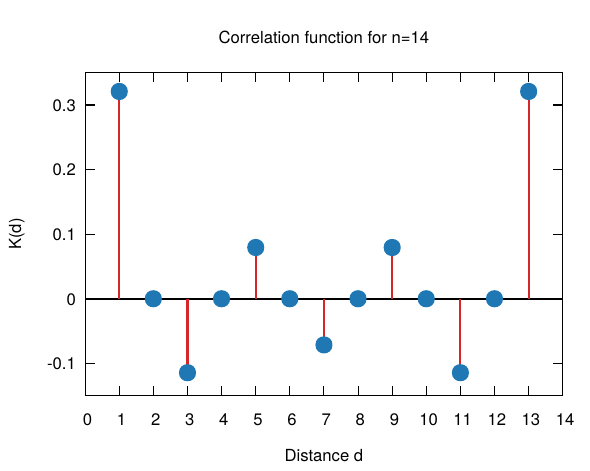}
\caption{
Two-point correlation function $K(d)$ as a function of the distance $d$ for a
cyclic system with $n=m=10$ (left figure) and $n=m=14$ (right figure) and
symmetric occupation of the Fourier modes. The most important feature is the
exact parity selection rule, according to which $K(d)$ vanishes identically for
all even lattice distances $d$. The analytic expression of $K(d)$ is reported
in Eq.~\ref{eq:kdhf}.
\label{fig:k_10_14}
} 
\end{figure}

\begin{figure}[h]
\centering
\includegraphics[width=0.45\linewidth]{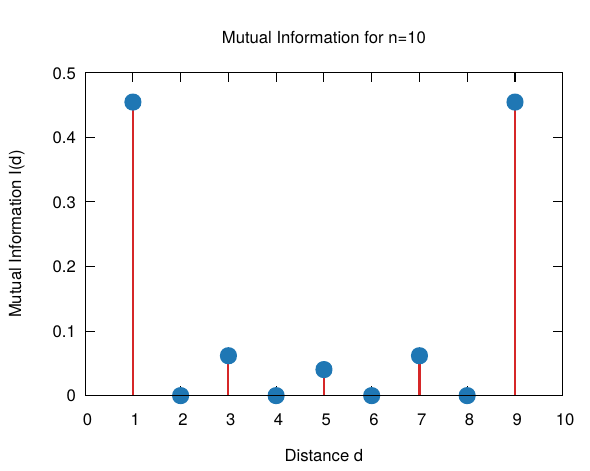}
\includegraphics[width=0.45\linewidth]{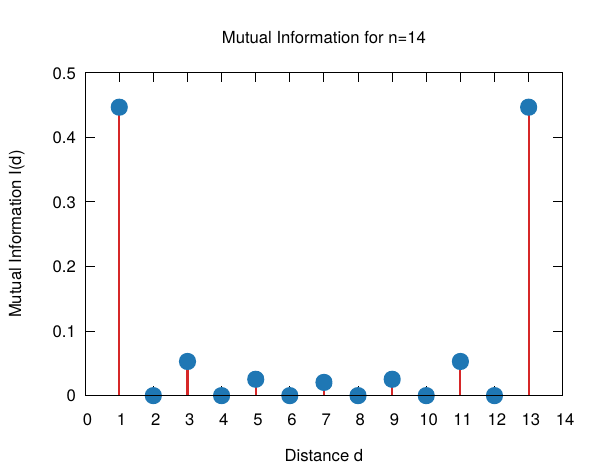}
\caption{
Mutual information $I(d)$ as a function of the distance $d$ for a cyclic system
with $n=m=10$ (left figure) and $n=m=14$ (right figure) at half filling.  The
most important feature is the exact vanishing of the mutual information at all
even lattice distances. This is the direct consequence of the parity selection
rule satisfied by the two-point correlator $K(d)$.  The analytic expression of
the mutual information is reported in Eq.~\ref{eq:mi-finale}.
 \label{fig:MI_10_14}
}
\end{figure}

The same analysis applies to larger cyclic systems with symmetric occupation of
Fourier modes in half filling. For $n=m=10$ and $n=m=14$, the occupied orbitals
again form a contiguous symmetric block around $\ell=0$, leading to the general
expression reported in Eq.~\ref{eq:kdhf}, with the parity selection rules $K(d)
= 0$ for all even  $d$. At the same time, the magnitude of $K(d)$ at odd
distances decreases with increasing system size, reflecting the spread of
electronic coherence over a larger number of sites.  The values of $K(d)$ for
these two cases are reported in Tab. \ref{tab:k-I} and shown in Fig.
\ref{fig:k_10_14}. The mutual information for these systems, obtained from Eq.
\ref{eq:mi-finale},  is reported in Tab. \ref{tab:k-I} and Fig.
\ref{fig:MI_10_14}.

\paragraph{Chemical interpretation.}

These results reveal a universal pattern of correlations in cyclic conjugated
hydrocarbons. While the molecular orbitals are fully delocalized over the
entire ring, the correlation structure in the localized basis is highly
structured and exhibits: \textit{i)} alternating behavior with strong
correlations between nearest neighbors; \textit{ii)} exact absence of
correlations between even-distance sites; \textit{iii)} weaker and oscillatory
correlations at longer distances.

From a chemical perspective, this shows that delocalization does not imply
uniform electron sharing across the molecule. Instead, the pattern of
correlations is governed by interference effects associated with the phase
structure of the molecular orbitals.

\subsection{Charged aromatic rings or general occupation patterns beyond half filling}

We now consider the more general situation in which the number of sites $m$
differs from the number of electrons $n$. From a chemical perspective, this
naturally occurs in charged aromatic systems, such as the cyclopentadienyl
anion ($n=5$, $n=6$), where  $\pi$ electrons occupy a closed-shell
configuration.  We keep using $N$ as the number of occupied orbitals per spin
sector.

As far as the two-point correlation function is concerned, the derivation
reported in Sec.~\ref{sec:kernel} remains valid and its analytic expression is
given in Eq.~\ref{eq:ker-general}.  $K(d)$ remains real and exhibits an
oscillatory structure determined by the interference of the occupied Fourier
modes.  Unlike the half filling case, the two-point correlator does not vanish
exactly at specific distances but it can display strong suppression at certain
distances.

Since mutual information $I(i,j)$ is a function of the eigenvalues of the
two-site reduced density matrix, which are fully determined by the correlation
matrix in Gaussian states, it inherits this behavior. The derivation reported
in Sec. \ref{sec:rdm-e-e} in its general form is valid and $I(i,j)$ is
expressed in Eq.~\ref{eq:mi-finale_general}.  In particular, one finds that the
strict selection rules of the half-filled case are replaced by a pattern of
oscillatory correlations with pronounced minima at even distances, but without
exact zeros.

Some examples of the values of $K(d)$ and of $I(d)$ are reported in
Tab.~\ref{tab:k-I} for the cases $n=6$ and $m=5,7$, $n=10$ and $m=9,11$, and
$n=14$ and $m=13,15$. The corresponding systems with half filling ($m=n$) are
also reported for comparison.

\begin{table*}
\caption{Values of the two-point correlation function $K(d)$ and of the corresponding mutual
information $I(d)$ for periodic one-dimensional fermionic systems with symmetric Fourier mode
occupation. The upper part of the table corresponds to $n=6$ electrons ($N=3$ per spin sector)
for rings with $m=5,6,7$ sites, the central part corresponds to $n=10$ electrons ($N=5$) for
$m=9,10,11$, while the lower part corresponds to $n=14$ electrons ($N=7$) for 
$m=13,14,15$. In the half-filled case ($m=n$), the two-point correlation function and the mutual information vanish exactly at even 
lattice distances, reflecting the bipartite structure of the periodic chain and the destructive 
interference generated by the symmetric occupation of the Fourier modes. Away from half filling
($m\neq n$), finite correlations are recovered at all distances.
\label{tab:k-I}}
\begin{center}
\begin{tabular}{|l|cc|cc|cc|cc|cc|cc|cc|}
\hline
      & \multicolumn{6}{|c|}{$n=6$} & & & & & & & &\\
\cline{2-7} 
      & $K(1)$ & $I(1)$ & $K(2)$ & $I(2)$ & $K(3)$ & $I(3)$ & & & & & & & & \\
\hline
$m=5$ & 0.32361 & 0.48664 & -0.12361 &  0.06448 & -0.12361 & 0.06448 & & & & & & & & \\
$m=6$ & 0.33333 & 0.48517 &  0       &  0       & -0.16667 & 0.11327 & & & & & & & & \\
$m=7$ & 0.32100 & 0.46159 &  0.07928 &  0.02579 & -0.11456 & 0.05412 & & & & & & & & \\
\hline
      & \multicolumn{10}{|c|}{$n=10$} & &  & & \\
\cline{2-11} 
      & $K(1)$ & $I(1)$ & $K(2)$ & $I(2)$ & $K(3)$ & $I(3)$ & $K(4)$ & $I(4)$ & $K(5)$ & $I(5)$ & & & & \\
\hline
$m=9$  & 0.31993 & 0.45211 & -0.05912 &  0.01419 & -0.11111 & 0.05045 &  0.07252 & 0.02138 & 0.07252 & 0.02138 & & & & \\
$m=10$ & 0.32361 & 0.45453 &  0       &  0       & -0.12361 & 0.06175 &  0       & 0       & 0.10000 & 0.04027 & & & & \\
$m=11$ & 0.31939 & 0.44748 &  0.04737 &  0.00907 & -0.10942 & 0.04870 & -0.05403 & 0.01180 & 0.06941 & 0.01950 & & & & \\
\hline
      & \multicolumn{14}{|c|}{$n=14$} \\
\cline{2-15} 
       & $K(1)$  & $I(1)$  & $K(2)$   & $I(2)$   & $K(3)$   & $I(3)$  & $K(4)$   & $I(4)$  & $K(5)$  & $I(5)$   & $K(6)$  & $I(6)$  & $K(7)$  & $I(7)$ \\
\hline
$m=13$ & 0.31909 & 0.44486 & -0.03961 &  0.00632 & -0.10846 & 0.04773 &  0.04344 & 0.00760 & 0.06771 & 0.01850 & -0.05138 & 0.01064 & -0.05138 & 0.01064 \\
$m=14$ & 0.32100 & 0.44654 &  0       &  0       & -0.11456 & 0.05297 &  0       & 0       & 0.07928 & 0.02525 &  0       & 0       & -0.07143 & 0.02048 \\
$m=15$ & 0.31889 & 0.44324 &  0.03408 &  0.00467 & -0.10787 & 0.04713 & -0.03645 & 0.00535 & 0.06667 & 0.01791 &  0.04120 & 0.00683 & -0.04982 & 0.00999 \\
\hline
\end{tabular}
\end{center}
\end{table*}

\subsection{Deviation from the Parity Selection Rule Away from Half Filling}

It is instructive to investigate how the parity selection rule is modified away from half filling. 
Let us introduce the deviation parameter
\begin{equation}
m=n+s=2N+s,
\end{equation}
where $s$ measures the distance from half filling. From
Eq.~\ref{eq:ker-general}, $K(d)$ for symmetric Fourier mode occupation and even
lattice distances, $d=2r$,  is
\begin{equation}
K(2r)
=
\frac{1}{m}
\frac{
\sin\left(\frac{2N\pi r}{m}\right)
}{
\sin\left(\frac{2\pi r}{m}\right)
}
\end{equation}
which can be rewritten as
\begin{equation}
\boxed{
K(2r)=\frac{(-1)^{r+1}}{m} \frac{\sin\left(\frac{\pi r s}{m}\right)}
{\sin\left(\frac{2\pi r}{m}\right)}
}
\label{eq:ker-s}
\end{equation}
showing how the parity selection rule breaks when $s\neq0$.

In particularly important cases $s=\pm1$, $K(d)$ becomes
\begin{equation}
K(2r)
=
\frac{(-1)^{r+1}}
{2m\cos\left(\frac{\pi r}{m}\right)},
\end{equation}
which explains the gradual increase of the even-distance correlations observed
numerically (see Tab.~\ref{tab:k-I}) as the distance increases. 
More generally, for arbitrary finite $s$, it follows from Eq.~\ref{eq:ker-s}
that the even-distance correlations may exhibit oscillatory behavior depending on
the value of $s$. The monotonic increase observed for $s=\pm1$ is therefore
not universal, although the overall scaling near half filling remains
controlled.

Indeed, for fixed distance \(2r\) and large system size $m$,
\begin{equation}
K(2r)
\simeq
\frac{(-1)^{r+1}s}{2m}.
\end{equation}
Thus, the violation of the parity selection rule scales as $1/m$ for fixed $s$.

From Eq.~\ref{eq:mi-finale} for the mutual information for Gaussian fermionic states, one
has that its small-$K$ expansion gives
\begin{equation}
I(d)
\propto
K(d)^2.
\end{equation}
Consequently,
\begin{equation}
I(2r)
=
O\left(\frac{s^2}{m^2}\right),
\end{equation}
showing that the parity selection rule is asymptotically recovered in the
thermodynamic limit whenever the deviation from half filling remains finite.

Therefore, the exact decoupling of sites belonging to the same sublattice at half filling
is not abruptly destroyed away from half filling, but instead evolves into a
finite-size effect whose magnitude is analytically controlled by the parameter
\(s=m-n\).

\section{Conclusions\label{sec:concl}}

In this work, we have developed a unified and analytical description of the
spatial structure of correlations in cyclic fermionic systems at the level of a
single Slater determinant with symmetric Fourier mode occupation.  The central
result of this work is that this structure is completely determined by the
occupied irreducible representations of the symmetry group. Once the occupied
Fourier modes are specified, the two-point correlator, reduced density
matrices, entanglement entropies and mutual information follow uniquely,
independently of the microscopic Hamiltonian generating those states.
Exploiting the cyclic symmetry of the system, we have shown that the
transformation from localized orbitals to molecular orbitals is simultaneously
a discrete Fourier transform, a Vandermonde matrix on the roots of unity, a
complex Hadamard matrix, and the character table of the cyclic group $C_m$.
This unified structure provides a transparent interpretation of the electronic
state in terms of symmetry-adapted modes and phase coherence.

Within this framework, the two-point correlation function is a truncated
Fourier sum over the occupied Fourier modes, which for symmetric Fourier mode
occupation patterns (corresponding to aromatic fillings) reduces to a simple
closed form, revealing the presence of exact selection rules in real space for
the special case of half filling. In particular, we have shown that in this
case the two-point correlation function vanishes identically for all even
distances along the ring.  This result reflects the bipartite structure of the
lattice and ultimately originates from the exact destructive interference
generated by the symmetric occupation of the Fourier modes.  We have further
demonstrated that this selection rule propagates to reduced density matrices
and to information-theoretic measures of correlation. In fermionic states
described by a single Slater determinant, the vanishing of $K(d)$ between two
sites implies a complete factorization of the two-body reduced density matrix
and, consequently, a zero mutual information. This establishes a direct link
between symmetry, interference, and the spatial organization of
correlations.\cite{Bogus12} The present results complement recent approaches
based on alternative quantum-information measures, such as spin concurrence,
which have also been shown to encode chemically relevant aspects of electronic
localization and bonding.\cite{Pitta26}

From a chemical perspective, these results provide a new interpretation of
delocalized electronic states in cyclic systems. While molecular orbitals are
fully delocalized over the ring, the underlying correlation structure is highly
non-uniform and exhibits exact nodes determined by symmetry. This highlights
the role of phase coherence and interference in shaping bonding patterns,
complementing traditional energy-based descriptions of aromaticity.

The framework developed here is completely independent of the specific form of
the Hamiltonian and applies to any system with cyclic symmetry described by a
single Slater determinant and symmetric occupation.  This applies equally to
H\"uckel models, tight-binding Hamiltonians, and mean-field solutions of more
general electronic Hamiltonians, provided that the resulting orbitals retain
the symmetry properties discussed above.  This shifts the focus from the
microscopic Hamiltonian to the symmetry and occupation structure of the
one-particle space, suggesting that many correlation patterns commonly
attributed to specific models are in fact universal consequences of the
occupied irreducible representations.  Deviations from this structure in
correlated wavefunctions are expected to provide direct signatures of genuine
many-body effects, a direction that will be explored in future work. This
perspective is consistent with previous quantum-information analyses of
molecular electronic structure, where orbital entanglement and mutual
information have been used to identify the most relevant correlation channels
and characterize electronic states.\cite{Barcza11,Bogus12} In more general
interacting systems, entanglement measures have proved capable of revealing
quantum phases and correlation effects that remain inaccessible to mean-field
descriptions.\cite{Chan23b}

Finally, we have shown that the framework extends naturally beyond half-filled
systems to charged cyclic structures with general occupation patterns. While
the Vandermonde structure of the wavefunction is preserved, the exact selection
rules are replaced by a more general interference pattern, demonstrating that
the correlation structure is fundamentally governed by symmetry and phase
coherence rather than by filling alone. The fact that these symmetry-induced
correlation patterns arise naturally in aromatic molecules highlights that the
present framework is not restricted to idealized lattice models, but applies
directly to experimentally accessible quantum systems. It therefore provides a
natural reference point for distinguishing symmetry-induced correlation
patterns from genuine many-body correlation effects in interacting systems.

It is also important to underline that the analysis reported here for cyclic
systems can be directly extended to periodic linear systems for which
translational symmetry mimics rotational symmetry. In this sense, cyclic
systems provide a particularly transparent realization of a more general
symmetry-based description of correlations in periodic fermionic systems.

\section*{Acknowledgments}
The author acknowledges funding from the Ministero dell'Universit\`a e della
Ricerca (MUR) under the Project PRIN 2022 number 2022W9W423.

\bibliography{jcp}

\end{document}